\documentstyle[prd,eqsecnum,aps,12pt]{revtex}

\begin{document}
\draft

\title{The Barton Expansion\\ 
and the\\
Path Integral Approach\\
in Thermal Field Theory}

\author{F. T. Brandt}
\address{Instituto de F\'\i sica, Universidade de S\~ao Paulo,
S\~ao Paulo, 05389-970 SP, Brazil\\
(fbrandt@if.usp.br)}

\author{D. G. C. McKeon}
\address{Dept. of Applied Mathematics, University of Western Ontario\\
London, Canada\\
(TMLEAFS@APMATHS.UWO.CA)}

\date{\today}

\maketitle
\vskip 1.0cm

\begin{abstract}
It has been shown how on-shell forward scattering amplitudes (the
``Barton expansion'') and quantum mechanical path integral (QMPI) can
both be used to compute temperature dependent effects in thermal field
theory. We demonstrate the equivalence of these two approaches and 
then
apply the QMPI to compute the high temperature expansion for the four-
point 
function in QED, obtaining results consistent with those previously
obtained from the Barton expansion.
\end{abstract}

\pacs{11.10.Wx}

\section{Introduction}\label{intro}
Temperature dependent effects can be computed in thermal field theory
in a variety of ways. In the real-time formalism, directly resorting 
to the
Feynman diagram approach results in awkward sums over so-called 
Matsubara
frequencies \cite{ref1}. It is possible to perform an analytic 
continuation in the time variable in this formalism to imaginary time,
and to then convert the sum over Matsubara frequencies into a contour 
integral.
The temperature dependent effects in one-loop processes are thereby
related to an on-shell forward scattering amplitude weighted by a 
statistical
factor. This is the so-called ``Barton expansion'' \cite{ref2,ref3}. 
An alternate
approach involves regulating the one-loop generating functional using 
the
zeta function \cite{ref4,ref5} and then computing matrix elements of 
the
form $<x|{\rm e}^{-H t}|y>$ 
that arise in this procedure by a quantum mechanical
path integral (QMPI). The QMPI is over a space compactfied in the 
temporal
direction; in place of a sum over Matsubara frequencies, a sum over 
winding
numbers is encountered \cite{ref6}.

These two approaches to computing radiative effects in thermal field 
theory
(the Barton expansion and the QMPI) are apparently quite disparate 
and there
is no obvious connection between the two techniques. In this paper, we
demonstrate that in fact they are equivalent. A calculation of the 
one-loop
amplitude with four external photons has been computed in thermal QED 
using
the Barton expansion \cite{ref7}; in this paper we reconsider this
computation using the QMPI and demonstrate that the results of these
approaches are compatible.

\section{The Barton Expansion and the QMPI}

Let us initially consider a model in which the term in the action 
bilinear in
the quantum field $\Phi$ is of the form.
\begin{equation}\label{eq1}
{\cal L}^{(2)}=-\Phi\left[\frac 1 2 \left(p-A\right)^2 + V\right]\Phi,
\end{equation}
where $p=-i\,\partial$ and $A_\mu$ and $V$ are functions of the 
external
fields. The one-loop generating functional is hence given by
\begin{equation}\label{eq2}
\Gamma^{(1)}\left[A_\mu,V\right]=\log {\rm sdet}^{-1/2}\left[
\frac1 2 \left(p-A\right)^2+V\right].
\end{equation}
In Euclidean space, a representation of $\Gamma^{(1)}$ is provided
by Schwinger's proper-time integral \cite{ref8}
\begin{equation}\label{eq3}
\Gamma^{(1)}\left[A_\mu,V\right]=\frac 1 2 
\int_0^\infty\frac {{\rm d}t} {t} {\rm Str}\;
{\rm e}^{- H t}
\end{equation}
($H\equiv \frac 1 2\left(p-A\right)^2+V$), or, upon regulating 
$\Gamma^{(1)}$ using the $\zeta$-function \cite{ref4,ref5},
\begin{equation}\label{eq4}
\begin{array}{ll}
\Gamma^{(1)}\left[A_\mu,V\right] & = \left.
\displaystyle{\frac {d} {ds}}\right 
|_{s=0} 
\left(\displaystyle{\frac 1 2} 
\displaystyle{\frac {1}{\Gamma\left(s\right)}} 
\displaystyle{\int_0^\infty} 
{\rm d}t t^{s-1} 
{\rm Str} \; {\rm e}^{- H t}\right)\\
{}&\equiv  \frac 1 2 \zeta'\left(0\right).
\end{array}
\end{equation}

An expansion due to Schwinger \cite{ref8}
\begin{equation}\label{eq5}
\begin{array}{ll}
{\rm Str}\;{\rm e}^{A+B} = {\rm Str}\;&\left[
\displaystyle{\int_0^1} {\rm d}\alpha_1
\;\delta\left(1-\alpha_1\right){\rm e}^{\alpha_1 A}\right.+\\
{}&\displaystyle{\int_0^1} 
{\rm d}\alpha_1\,{\rm d}\alpha_2\;\delta\left(1-\alpha_1-
\alpha_2\right)
{\rm e}^{\alpha_1 A}B{\rm e}^{\alpha_2 A} +\\
{}&\left.\displaystyle{\int_0^1} 
{\rm d}\alpha_1\,{\rm d}\alpha_2\,{\rm d}\alpha_3
\;\delta\left(1-\alpha_1-\alpha_2-\alpha_3\right)
{\rm e}^{\alpha_1 A}B{\rm e}^{\alpha_2 A}B{\rm e}^{\alpha_3 
A}+\cdots\right]
\end{array},
\end{equation}
can be used in (\ref{eq3}) or (\ref{eq4}) to extract contributions to
$\Gamma^{(1)}$ to any order in $A_\mu$ and $V$, thereby allowing us to
compare one-loop Green's functions. The Str operation, if performed 
in 
momentum space, results in a loop-momentum integral while integrals 
over
the paremeters $\alpha_i$ are direct analogues of Feynman parameters
integrals. In fact, from (\ref{eq3}), the expression for the $N$-point
function is identical to that obtained from the Feynman diagram 
approach
once the denominators from individual propagators are combined. 
However, we
note that the regulated expression for $N$-point function obtained by 
using
the Schwinger expansion of (\ref{eq5}) in conjunction with the 
regulated
one-loop generating functional of (\ref{eq4}) cannot be obtained by 
evaluating
Feynman diagrams using an action that has been regulated in some way;
the functional determinant in (\ref{eq2}) is what has been regulated 
and
hence the designation ``operator regularization'' is used in 
\cite{ref5}.

In thermal field theory, fields are periodic in Euclidean time if they
are Bosonic and antiperiodic if they are Fermionic; hence time-like
momentum variables in these two cases are of the form 
$n\left(2\pi/\beta\right)$ and 
$\left(n+1/2\right)\left(2\pi/\beta\right)$
respectively \cite{ref1}. A sum over $n$ (the ``Matsubara sum'') 
replaces
the usual integral over the time-like component of the loop-momentum. 
It is
possible to convert this sum into a contour integral \cite{ref10} and 
then
to extract the temperature dependence ($T\equiv1/\beta$) by 
considering
a forward on-shell elastic scattering amplitude with an appropriate 
statistical weighting factor. To illustrate this, let us consider the
particular case where in (\ref{eq1}), $A_\mu=0$ and $V=\frac 1 2 
F\left(x\right)$,
a background scalar field in $D$ dimensions. From (\ref{eq3}) then, 
\begin{equation}\label{eq6}
\Gamma^{(1)}\left[F\right]=\frac 1 2 \int_0^\infty\frac{{\rm d}t}{t}
{\rm tr}\;{\rm e}^{-\frac 1 2\left(p^2+F\right)t},
\end{equation}
which to second order in $F$ gives, using (\ref{eq5})
\begin{equation}\label{eq7}
\simeq\frac 1 2 \int_0^\infty\frac{{\rm d}t}{t} {\rm tr}\;
\left(-t\right)^2\int_0^1 {\rm d}u
{\rm e}^{-\left(1-u\right)p^2 t}F
{\rm e}^{-u p^2 t}F.
\end{equation}
Computing the functional trace in (\ref{eq7}) in momentum space as in
\cite{ref8,ref5} we find after integrating over $t$ that
\begin{equation}\label{eq8}
=\frac 1 2\int{\rm d}^D p \int \frac{{\rm d}^D q}{\left(2\pi\right)^D}
\int_0^1{\rm d} u \frac{\tilde F\left(p\right)\tilde F\left(-p\right)}
{\left[\left(1-u\right)\left(p+q\right)^2+u q^2\right]^2}.
\end{equation}
In (\ref{eq8}) we see that upon integrating over $u$ the Feynman 
integral
for the contributions of the two-point functions to the one-loop 
generating
functional is recovered,
\begin{equation}\label{eq9}
=\frac 1 2\int{\rm d}^D p \tilde F\left(p\right)\tilde F\left(-
p\right)
\int \frac{{\rm d}^D q}{\left(2\pi\right)^D}
\frac{1} {q^2\left(q+p\right)^2}.
\end{equation}
In thermal field theory this becomes
\begin{equation}\label{eq10}
\begin{array}{ll}
\displaystyle{= \frac 1 2} \int{\rm d}^{D-1}p\sum_m
& \tilde  F\left(\vec p,\frac{2\pi m}{\beta}\right)
 \tilde  F\left(-\vec p,-\frac{2\pi m}{\beta}\right)
 \left(
\displaystyle{
\frac{1}{\beta}}
\;\sum_n\int
\displaystyle{
\frac{{\rm d}^{D-1}q}{\left(2\pi\right)^{D-1}}}
\right.\\
&\left.
\displaystyle{
 \frac{1}{\vec q^2+\left(\frac{2\pi m}{\beta}\right)^2}}
\displaystyle{
\frac{1}{\left(\vec q+\vec p\right)^2+\left(\frac{2\pi}{\beta}
\left(n+m\right)\right)^2}}\right)
\end{array}.
\end{equation}

The sum over $n$ in (\ref{eq10}) can be converted into a contour 
integral
using the formula \cite{ref9}
\begin{equation}\label{eq11}
\begin{array}{ll}
\displaystyle{
\frac{2\pi i}{\beta}}
\sum_{n=-\infty}^{n=\infty} f\left(\frac{2\pi i n}{\beta}\right)=&
\displaystyle{\frac 1 2} 
\int_{-\infty}^{\infty}{\rm d} z \left[f\left(i z\right)+
f\left(-i z\right)\right]\\
&+
\displaystyle{\int}_{-\infty+\delta}^{\infty+\delta}{\rm d} z 
\left[\displaystyle{\frac{f\left(i z\right)+
f\left(-i z\right)}{{\rm e}^{i\beta z}-1}}\right]
\end{array},
\end{equation}
provided $f\left(i z\right)$ has no singularities along the imaginary
$z$ axis. All the temperature dependence in (\ref{eq11}) resides in 
the
second term; hence we find that the temperature dependence in
(\ref{eq10}) comes from
\begin{equation}\label{eq12}
\begin{array}{ccc}
{}&
\displaystyle{\int}
\displaystyle{\frac{{\rm d}^{D-1} q}{\left(2 \pi\right)^{D-1}}}
\displaystyle{\frac{1}{2\pi i}} \int_{-
i\infty+\delta}^{i\infty+\delta}
N\left(z\right)
\left[
\displaystyle{\frac{1}{\left(\vec q^2-z^2\right)}}\;\;
\displaystyle{
\frac{1}{\left(\vec p+\vec q\right)^2-\left(p_0+z\right)^2}}\right]
&{}\\
{}&{}&{}\\
{}&\left(N\left(z\right)\equiv\left({\rm e}^{\beta z}-1\right)^{-
1}\right)
&{}
\end{array}
\end{equation}
Performing a partial fraction expansion in (\ref{eq12}) we obtain
\begin{equation}\label{eq13}
\begin{array}{ll}
=\displaystyle{\int}\displaystyle{\frac{{\rm d}^{D-1} q}
{\left(2\pi\right)^{D-1}}}
\displaystyle{\frac{1}{2\pi i}}
\displaystyle{\int_{-i \infty+\delta}^{i \infty+\delta}}{\rm d} z 
&
N\left(z\right)\left[\displaystyle{\frac{1}{2 |\vec q|}}\left(
\displaystyle{\frac{1}{z-|\vec q|}}-
\displaystyle{\frac{1}{z+|\vec q|}}\right)\right]\\
{}&{}\\
{}&
\left[\displaystyle{\frac{1}{2 |\vec p+\vec q|}}\left(
\displaystyle{\frac{1}{z+p_0-|\vec p+\vec q|}}-
\displaystyle{\frac{1}{z+p_0+|\vec p+\vec q|}}\right)\right]
\end{array}.
\end{equation}
Since $p_0=2 \pi i m/\beta$, we see that 
$N\left(z\right)=N\left(z+p_0\right)$.
This permits us to close the contour integral in $z$ on the right 
half plane,
and then to use the residue theorem to obtain
\begin{equation}\label{eq14}
=\int\frac{{\rm d}^{D-1} q}{\left(2 \pi\right)^{D-1}}
\frac{N\left(|\vec q|\right)}{2 |\vec q|}
\left.
\left[\frac{1}{\left(q+p\right)^2}+
      \frac{1}{\left(q-p\right)^2}\right]\right|_{q_0=|\vec q|}
\end{equation}
This is the so-called ``Barton amplitude'' \cite{ref2,ref3}; viz an
expression the on-shell forward scattering amplitude with two external
fields, weighted by the statistical factor $N\left(|\vec q|\right)$. 
This
can be generalized to handle more external legs, vertices with 
momentum
dependence, and internal particles with Fermi statistics \cite{ref3}. 
In
particular, the non-linear four-point interaction of photons has been 
analysed using this approach \cite{ref7}

An alternate procedure for examining thermal Green's functions is to 
use the
QMPI. In this approach, we use the fact that a path integral can be 
used
to represent the matrix element in (\ref{eq3}) and (\ref{eq4}),
\begin{equation}\label{eq15}
\begin{array}{ccc}
{}&
<x|exp-\left[
\displaystyle{\frac 1 2} \left(p-A\right)^2+V\right]\,t|y>=
&{}\\
{}&
{\rm P} \!\!\!\!\!\:\! {\rm I}\;\;
\displaystyle{
\int_{\!\!\!\!\!\!\!\!\!{_{{_{\stackrel{q\left(0\right)=y}{\rm 
per}}}}}}
^{\!\!\!\!\!\!\!\!{^{^{q\left(t\right)=x}}}}}\!\!\!\!
{\cal D} q\left(\tau\right)
{\rm exp}\left\{\displaystyle{\int}_0^t {\rm d} \tau 
\left[\displaystyle{-\frac 1 2} \dot{q}^2\left(\tau\right)+
i A\left(q\left(\tau\right)\right)\cdot q\left(\tau\right) - 
V\left(q\left(\tau\right)\right)\right]\right\}&{}\\
{}&
\left({\rm P} \!\!\!\!\!\:\! {\rm I}\;\; 
{\rm is\;the\;path\;ordering}\right)&{}
\end{array},
\end{equation}
where ``per'' means invariance under the replacement
$q_0\left(\tau\right)\rightarrow q_0\left(\tau\right)+\beta$.

As is discussed in \cite{ref10,ref6} (\ref{eq15}) can be rewritten as
\begin{equation}\label{eq16}
\sum_{n=-\infty}^{n=\infty}{\rm e}^{i \delta_n}
{\rm P} \!\!\!\!\!\:\! {\rm I}\;\;
\int_{_{_{\!\!\!\!\!\!\!{_{_{_{q\left(0\right)=y}}}}}}}
^{\!\!\!\!\!\!\!\!\!{^{^{q\left(t\right)=\left(x_0+n\beta,\vec 
x\right)}}}}
\!\!\!\!\!\!\!\!\!\!\!\!\!\!\!\!\!\!\!\!\!{\cal D} q\left(\tau\right)
{\rm exp}\left\{\int_0^t {\rm d} \tau 
\left[\displaystyle{-\frac 1 2} \dot{q}^2\left(\tau\right)+
i A\left(q\left(\tau\right)\right)\cdot q\left(\tau\right) - 
V\left(q\left(\tau\right)\right)\right]\right\},
\end{equation}
with $\delta_n=0$ for Bosons and $\delta_n=n \pi$ for Fermions.

We will restrict ourselves to examining (\ref{eq16}) to zeroth order 
in $V$
and second order in $A_\mu$ (higher orders can be treated in a 
similar 
fashion). If
\begin{equation}\label{eq17}
\left(2\pi\right)^{\frac{D-1}{2}}\sqrt{\beta} 
A_\mu\left(q\left(\tau_i\right)\right)=\epsilon_{i_\mu}
{\rm exp}-\left(i k_i \cdot q\left(\tau_i\right)\right),
\end{equation}
then we have
\begin{equation}\label{eq18}
=\int_0^t{\rm d} \tau_1\int_0^{\tau_1}{\rm d} \tau_2
\sum_{n=-\infty}^{\infty}{\rm e}^{i \delta_n}
\frac{1}{\left(2\pi\right)^{D-1}\beta}
\int_{_{_{\!\!\!\!\!\!\!{_{_{_{q\left(0\right)=y}}}}}}}
^{\!\!\!\!\!\!\!\!\!{^{^{q\left(t\right)=\left(x_0+n\beta,\vec 
x\right)}}}}
\!\!\!\!\!\!\!\!\!\!\!\!\!\!\!\!\!\!\!\!\!{\cal D} q\left(\tau\right)
{\rm exp}\left\{\int_0^t {\rm d} \tau
\left[-\frac 1 2 \dot{q}^2\left(\tau\right)-
\gamma\left(\tau\right)\cdot q\left(\tau\right)\right]\right\},
\end{equation}
where 
$\gamma\left(\tau\right)=i\sum_{i=1}^{2}\left[\delta\left(\tau-
\tau_i\right)
k_i+\delta\left(\dot{\tau}-\tau_i\right)\epsilon_i\right]$
provided we keep only terms linear in $\epsilon_i$.

Evaluating this path integral and extracting a high temperature 
expansion
for the two-point function from it is outlined in \cite{ref11}. 
However,
it is not apparent how the results one obtains from this approach are 
at
all related to what is extracted from the Schwinger expansion of 
(\ref{eq5}) in conjunction with (\ref{eq3}), and hence the Barton 
expansion. Indeed, the Schwinger expansion of the left side of 
(\ref{eq15})
to second order in $A$ is
\begin{equation}\label{eq19}
\begin{array}{lrl}
I_{x y}=<x|
&\left[
\displaystyle{\int_0^1} {\rm d}u\;{\rm e}^{-\frac 1 2\left(1-
u\right)p^2 t}
\left(\displaystyle{-\frac{t A^2}{2}}\right)
{\rm e}^{-\frac 1 2 u p^2 t}\right. 
&\\
&
+\displaystyle{\int_0^1}
{\rm d}u\;u\;{\rm e}^{-\frac 1 2 \left(1-u\right) p^2 t}
&\left(\displaystyle{\frac 1 2}\right)
\left(p\cdot A + A\cdot p\right) t
{\rm e}^{-\frac 1 2 u\left(1-v\right)p^2 t}
\\
&
&
\left.\!\!\!\!\!\!\!\!\!\!\!\!\!\!\times
\left(\displaystyle{\frac 1 2}\right)
\left(p\cdot A+A\cdot p\right)t
{\rm e}^{-\frac 1 2 u v p^2 t} \right]|y>\; ;
\end{array}
\end{equation} 
we can also expand (\ref{eq16}) to second order in $A$ to obtain
\begin{equation}\label{eq20}
\begin{array}{ll}
J_{xy}=
\displaystyle{\int_0^t}{\rm d}\tau_1
\displaystyle{\int_0^{\tau_1}}{\rm d}\tau_2
\sum_{n=-\infty}^{\infty}{\rm e}^{i\delta_n}
\displaystyle{
\int_{_{_{\!\!\!\!\!\!\!{_{_{_{q\left(0\right)=y}}}}}}}^{\!\!\!\!\!\!
\!\!\!{^{^{q\left(t\right)=\left(x_0+n\beta,\vec x\right)}}}}}
\!\!\!\!\!\!\!\!\!\!\!\!\!\!\!\!\!\!\!\!\!
&{\cal D} q\left(\tau\right)
{\rm exp}
\displaystyle{\int_0^t} {\rm d} \tau 
\left(\displaystyle{-\frac 1 2} \dot{q}^2\left(\tau\right)\right)\\
&
\left[i\;\dot{q}\left(\tau_1\right)\cdot 
A\left(q\left(\tau_1\right)\right)
\right]
\left[i\;\dot{q}\left(\tau_2\right)\cdot 
A\left(q\left(\tau_2\right)\right)
\right]
\end{array}.
\end{equation}
In order to demonstrate the equivalence of $I_{xy}$ and $J_{xy}$,
we split up the path integral in (\ref{eq20}) so that
\begin{equation}\label{eq21}
\begin{array}{c}
J_{xy}=
\displaystyle{\int_0^t}{\rm d}\tau_1
\displaystyle{\int_0^{\tau_1}}{\rm d}\tau_2
\displaystyle{\sum_{n=-\infty}^{\infty}}
{\rm e}^{i\delta_n}
\displaystyle{\int{\rm d}^Db_1}\displaystyle{\int{\rm d}^Db_2}
\displaystyle{
\int_{_{_{\!\!\!\!\!\!\!{_{_{q\left(0\right)=y}}}}}}^{\!\!\!\!\!\!
\!\!\!{^{^{^{q\left(\tau_2\right)=b_2}}}}}}
\!\!\!\!\!\!\!\!\!\!\!\!\!\!\!\!\!\!\!\!\!
\;\;\;\;\;\;\;\;\;
{\cal D} q\left(\tau\right)
\displaystyle{
\int_{_{\!\!\!\!\!\!\!{_{_{_{q\left(\tau_2\right)=b_2}}}}}}
^{\!\!\!\!\!\!\!\!\!{^{^{^{q\left(\tau_1\right)=b_1}}}}}}
\!\!\!\!\!\!\!\!\!\!\!\!\!\!\!\!\!\!\!\!\!
\;\;\;\;\;\;\;\;\;
{\cal D} q\left(\tau\right)
\displaystyle{
\int_{_{\!\!\!\!\!\!\!{_{_{_{q\left(\tau_1\right)=b_1}}}}}}
^{\!\!\!\!\!\!\!\!\!{^{^{^{q\left(t\right)=\left(x_0+n\beta,\vec 
x\right)
}}}}}}
\!\!\!\!\!\!\!\!\!\!\!\!\!\!\!\!\!\!\!\!\!
{\cal D} q\left(\tau\right) \\
{}{}\\
\times
{\rm e}^{-\frac 1 2 \left[
\int_0^{\tau_2}{\rm d}\tau+
\int_{\tau_2}^{\tau_1}{\rm d}\tau+
\int_{\tau_1}^t{\rm d}\tau\right]
\dot{q}^2\left(\tau\right)}
\left[i\left(\displaystyle{\frac{q\left(\tau_1+\epsilon_1\right)-
q\left(\tau_1-\epsilon_1\right)}{2\epsilon_1}}\right)\cdot 
A\left(b_1\right)\right]\\
{}{}\\
\times
\left[i\left(\displaystyle{\frac{q\left(\tau_2+\epsilon_2\right)-
q\left(\tau_2-\epsilon_2\right)}{2\epsilon_2}}\right)\cdot 
A\left(b_2\right)\right]
\end{array}.
\end{equation}
In (\ref{eq21}) we have replaced the time derivative 
$\dot{q}\left(\tau_i\right)$ by 
$\left(q\left(\tau_i+\epsilon_i\right)-
q\left(\tau_i-\epsilon_i\right)/2\epsilon_i\right)$,
which is valid in the limit $\epsilon_i\rightarrow 0$.

The path integrals in (\ref{eq21}) can be evaluated using the result
\begin{equation}\label{eq22}
\begin{array}{c}
\displaystyle{
\int_{_{\!\!\!\!\!\!\!{_{_{_{q\left(\tau_2\right)=b_2}}}}}}
^{\!\!\!\!\!\!\!\!\!{^{^{^{q\left(\tau_1\right)=b_1}}}}}}
\!\!\!\!\!\!\!\!\!
{\cal D} q\left(\tau\right){\rm exp}
\displaystyle{\int_{\tau_2}^{\tau_1}}{\rm d}\tau
\left(-\displaystyle{\frac 1 2}\dot{q}^2\left(\tau\right)-
\gamma\left(\tau\right)\cdot q\left(\tau\right)\right)
\\
\\
=\displaystyle{\frac{1}{\left[2\pi\left(\tau_1-\tau_2\right)\right]^
{\frac D 2}}}
{\rm exp}\left\{-\displaystyle{\frac{\left(b_1-b_2\right)^2}
{2\left(\tau_1-\tau_2\right)}}-\displaystyle{\frac{1}{\tau_1-\tau_2}}
\displaystyle{\int_{_{_{\!\!\!\tau_2}}}^{^{\!\tau_1}}}
\!\!\!{\rm d}\tau
\left[b_1\left(\tau-\tau_2\right)+b_2\left(\tau_1-\tau\right)\right]
\cdot\gamma\left(\tau\right)\right.
\\
\\
\left.
\displaystyle{-\frac 1 2}\displaystyle{\int_{\tau_2}^{\tau_1}}
{\rm d}\tau {\rm d}\tau' G\left(\tau,\tau';\tau_2,\tau_1\right)
\gamma\left(\tau\right)\cdot\gamma\left(\tau'\right)\right\},
\end{array}
\end{equation}
where 
\begin{equation}\label{eq23}
G\left(\tau,\tau';\tau_2,\tau_1\right)\equiv
\frac 1 2 |\tau-\tau'|-\frac{\tau_1+\tau_2}{2\left(\tau_1-
\tau_2\right)}
\left(\tau+\tau'\right)+\frac{\tau\tau'}{\tau_1-\tau_2}+
\frac{\tau_1\tau_2}{\tau_1-\tau_2}.
\end{equation} 

By systematically using (\ref{eq22}), (\ref{eq21}) becomes
\begin{equation}\label{eq24}
\begin{array}{c}
J_{x y}=-\displaystyle{\int_0^       t}{\rm d}\tau_1
         \displaystyle{\int_0^{\tau_1}}{\rm d}\tau_2
         \displaystyle{\int}{\rm d}b_1{\rm d}b_2
       \;\displaystyle{\sum_n}{\rm e}^{i\delta_n}
\left[\displaystyle{\frac{1}{2\pi\left(t-\tau_1\right)}}
      \displaystyle{\frac{1}{2\pi\left(\tau_1-\tau_2\right)}}
      \displaystyle{\frac{1}{2\pi\tau_2}}\right]^{\frac D 2}
      A_{\mu_1}\left(b_1\right) A_{\mu_2}\left(b_2\right)
\\
\\
\times
{\rm exp}-\displaystyle{\frac 1 2}\left[
\displaystyle{\frac{\left(x^*-b_1\right)^2}{t-\tau_1}}+
\displaystyle{\frac{\left(b_1-b_2\right)^2}{\tau_1-\tau_2}}+
\displaystyle{\frac{\left(b_2-y\right)^2}{\tau_2}}\right]
\\
\\
\times
\left\{\displaystyle{\frac{\delta_{\mu_1\mu_2}}{4\epsilon_1\epsilon_2}}
G\left(\tau_2+\epsilon_2,\tau_1-\epsilon_1;\tau_2,\tau_1\right)+
\displaystyle{\frac 1 4}
\left[
\displaystyle{\frac{x^*-b_1}{t-\tau_1}}+
\displaystyle{\frac{b_1-b_2}{\tau_1-\tau_2}}\right]_{\mu_1}
\left[
\displaystyle{\frac{b_1-b_2}{\tau_1-\tau_2}}+
\displaystyle{\frac{b_2-y}{\tau_2}}\right]_{\mu_2}\right\}
\end{array}.
\end{equation}
(we have defined $x^*\equiv\left(x^0+n\beta,\vec x\right)$).
Using the Fourier transform
\begin{equation}\label{eq25}
\frac{{\rm e}^{-x^2/2t}}{\left(2\pi t\right)^{D/2}}=
\int \frac{{\rm d}^Dp}{\left(2\pi\right)^D}\;
{\rm e}^{-i p \cdot x - \frac 1 2 p^2 t},
\end{equation}
we can rewrite (\ref{eq24}) in the form
\begin{equation}\label{eq26}
\begin{array}{c}
      = -\displaystyle{\int_0^       t}{\rm d}\tau_1
         \displaystyle{\int_0^{\tau_1}}{\rm d}\tau_2
         \displaystyle{\int}{\rm d}b_1\;{\rm d}b_2
A_{\mu_1}\left(b_1\right)\,A_{\mu_2}\left(b_2\right)
\displaystyle{\int}
\displaystyle{\frac{{\rm d}p_1}{\left(2\pi\right)^D}}
\displaystyle{\frac{{\rm d}p_2}{\left(2\pi\right)^D}}
\displaystyle{\frac{{\rm d}p_3}{\left(2\pi\right)^D}}
{\rm e}^{-\frac 1 2\left[p_1^2\left(t-\tau_1\right)+
p_2^2\left(\tau_1-\tau_2\right)+p_3^2\tau_2\right]}
\\
\\
\times
\;
\displaystyle{\sum_n} {\rm e}^{i\delta_n}
\left\{
\displaystyle{\frac{\delta_{\mu_1 \mu_2}}{4 \epsilon_1\epsilon_2}}
G\left(\tau_2+\epsilon_2,\tau_1-\epsilon_1;\tau_2,\tau_1\right)+
\displaystyle{\frac 1 4}
\left[
\displaystyle{\frac{1}{t-\tau_1}}
\left(i\displaystyle{\frac{\partial}{\partial p_1}}\right)+
\displaystyle{\frac{1}{\tau_1-\tau_2}}
\left(i\displaystyle{\frac{\partial}{\partial 
p_2}}\right)\right]_{\mu_1}
\right.
\\
\\
\left.
\times
\left[
\displaystyle{\frac{1}{\tau_1-\tau_2}}
\left(i\displaystyle{\frac{\partial}{\partial p_2}}\right)+
\displaystyle{\frac{1}{\tau_2}}
\left(i\displaystyle{\frac{\partial}{\partial 
p_3}}\right)\right]_{\mu_2}
\right\}
{\rm e}^{-\left[p_1\cdot\left(x^*-b_1\right)+
                p_2\cdot\left(b_1-b_2\right)+
                p_3\cdot\left(b_2-y\right)\right]}
\end{array}
\end{equation}
Integrating by parts with respect to the derivatives occurring in 
(\ref{eq26}) leaves us with
\begin{equation}\label{eq27}
\begin{array}{c}
      = -\displaystyle{\frac 1 4}
         \displaystyle{\int_0^       t}{\rm d}\tau_1
         \displaystyle{\int_0^{\tau_1}}{\rm d}\tau_2
         \displaystyle{\int}
         \displaystyle{\frac{{\rm d}p_1\;{\rm d}p_2\;{\rm d}p_3}
                            {\left(2\pi\right)^D}}
\;
\displaystyle{\sum_n} {\rm e}^{i\delta_n}\;
{\rm e}^{-i p_1\cdot x^* + i p_3\cdot y}
\displaystyle{\int}
\displaystyle{\frac{{\rm d}b_1}{\left(2\pi\right)^D}}
{\rm e}^{i\left(p_1-p_2\right)\cdot b_1}
A_{\mu_1}\left(b_1\right)
\\
\\
\times
\displaystyle{\int}
\displaystyle{\frac{{\rm d}b_2}{\left(2\pi\right)^D}}
{\rm e}^{i\left(p_2-p_3\right)\cdot b_2}
A_{\mu_2}\left(b_2\right)
\left\{\delta_{\mu_1\mu_2}\left[
\displaystyle{
\frac{G\left(\tau_2+\epsilon_2,\tau_1-\epsilon_1;\tau_2,\tau_1\right)}
{\epsilon_1\epsilon_2}}+
\displaystyle{\frac{1}{\tau_1-\tau_2}}\right]\right.
\\
\\
\left.
-\left(p_1+p_2\right)_{\mu_1}\, \left(p_1+p_3\right)_{\mu_2}\right\}
\displaystyle{
{\rm e}^{-\frac t 2\left[p_1^2\left(1-\sigma_1\right)+
                          p_2^2\left(\sigma_1-\sigma_2\right)+
                          p_3^2 \sigma_2\right]}}
\end{array}
\end{equation}
From the definition of $G$ in (\ref{eq23}), we see that
\begin{equation}\label{eq28}
G\left(\tau_2+\epsilon_2,\tau_1-\epsilon_1;\tau_2,\tau_1\right)=
\frac 1 2|\tau_1-\tau_2-\epsilon_1-\epsilon_2|-
\frac 1 2\left(\tau_1-\tau_2-\epsilon_1-\epsilon_2\right)
-\frac{\epsilon_1\epsilon_2}{\tau_1-\tau_2}
\end{equation}
So that in the limit $\epsilon_1,\;\epsilon_2\rightarrow 0$, 
(\ref{eq27}) reduces to
\begin{equation}\label{eq29}
\begin{array}{c}
      = \displaystyle{\frac {t^2} {4}}
         \displaystyle{\int_0^       1}{\rm d}\sigma_1
         \displaystyle{\int_0^{\sigma_1}}{\rm d}\sigma_2
         \displaystyle{\int}
         \displaystyle{\frac{{\rm d}p_1\;{\rm d}p_2\;{\rm d}p_3}
                            {\left(2\pi\right)^D}}
\;
\displaystyle{\sum_n} {\rm e}^{i\delta_n}\;
{\rm e}^{-i p_1\cdot x^* + i p_3\cdot y}
\tilde A_{\mu_1}\left(p_1-p_2\right)\,
\tilde A_{\mu_2}\left(p_2-p_3\right)
\\
\\
\times
\left\{-\delta_{\mu_1\mu_2}
\displaystyle{\frac{\delta\left(\sigma_1-\sigma_2\right)}{t}}+
\left(p_1+p_2\right)_{\mu_1}\, \left(p_2+p_3\right)_{\mu_2}\right\}
\displaystyle{
{\rm e}^{-\frac t 2\left[p_1^2\left(1-\sigma_1\right)+
                          p_2^2\left(\sigma_1-\sigma_2\right)+
                          p_3^2 \sigma_2\right]}}
\end{array},
\end{equation}
provided $\tau_i=\sigma_i t$ and $\tilde A_\mu$ is the Fourier 
transform of
$A_\mu$. The Poisson resumation formula
\begin{equation}\label{eq30}
\sum_{m=-\infty}^{\infty}f\left(m\right)=
\sum_{m=-\infty}^{\infty}\int_{-\infty}^{\infty}{\rm e}^{2\pi i m}
f\left(\mu\right)
\end{equation}
can now be used to show that if $\delta_n=n\pi\epsilon$
($\epsilon=0$ for Bosons and $\epsilon=1$ for Fermions) then
\begin{equation}\label{eq31}
\sum_{m=-\infty}^{\infty}{\rm e}^{i\delta_n}{\rm e}^{-i p^0_1 
\left(n\beta\right)}=
\sum_{m=-\infty}^{\infty}\frac{2\pi}{\beta}\delta\left(p^0_1-
\frac{2\pi\left(n+\epsilon/2\right)}
{\beta}\right).
\end{equation}
Substitution of (\ref{eq31}) into (\ref{eq29}) leads to
\begin{equation}\label{eq32}
\begin{array}{c}
J_{xy} = \displaystyle{\frac {t^2} {4}}
         \displaystyle{\int_0^       1}{\rm d}\sigma_1
         \displaystyle{\int_0^{\sigma_1}}{\rm d}\sigma_2
         \displaystyle{\int}
         \displaystyle{\frac{{\rm d}p_1\;{\rm d}p_2\;{\rm d}p_3}
                            {\left(2\pi\right)^D}}
\;
\displaystyle{\sum_n} \displaystyle{\frac{2\pi}{\beta}}
\delta\left(p^0_1-
\displaystyle{\frac{2\pi\left(n+\epsilon/2\right)}{\beta}}y
\right)
{\rm e}^{-i p_1\cdot x + i p_3\cdot y}
\\
\\
\times
\tilde A_{\mu_1}\left(p_1-p_2\right)\,
\tilde A_{\mu_2}\left(p_2-p_3\right)
\left\{-\delta_{\mu_1\mu_2}
\displaystyle{\frac{\delta\left(\sigma_1-\sigma_2\right)}{t}}+
\left(p_1+p_2\right)_{\mu_1}\, \left(p_2+p_3\right)_{\mu_2}\right\}
\\
\\
\times
\displaystyle{
{\rm e}^{-\frac t 2\left[p_1^2\left(1-\sigma_1\right)+
                          p_2^2\left(\sigma_1-\sigma_2\right)+
                          p_3^2 \sigma_2\right]}}
\end{array}.
\end{equation}
It is immediately apparent that (\ref{eq32}) is identical to
(\ref{eq19}) upon insertion of a complete set of momentum states into 
the
later, as in \cite{ref8,ref5}, provided these momentum states have 
Matsubara
frequencies associated with their time-like component. One can easily 
generalize
this result to show that one can obtain the same result for the 
matrix element
$M_{xy}=<x|exp-\left[\frac 1 2\left(p-A\right)^2+V\right]\, t|y>$
whether one uses the Schwinger expansion or the quantum mechanical 
path
integral for $T>0$; consequently the Barton expansion and the 
expansion of
the quantum mechanical path integral are equivalent procedures, 
despite
their apparent differences.

Let us now consider how the four-point function can be computed in
quantum electrodynamics \cite{ref12} and compare the result with that
obtained in \cite{ref7} where the Barton expansion was employed.

\section{The four point function}
The one-loop generating function for Green's functions with external 
photon
lines in quantum electrodynamics is
\begin{equation}\label{eq33}
i\Gamma^{(1)}\left[A_\mu\right]={\rm ln }\;{\rm det}^{1/2}
\frac 1 2\left[\left(p\!\!\!/ - A \!\!\!/\right)^2+m^2\right]
\end{equation}
If $\left\{\gamma_\mu,\gamma_\nu\right\}=2\eta_{\mu\nu}$ and
$\sigma_{\mu\nu}=\frac i 2 \left[\gamma_\mu,\gamma_\nu\right]$, then
\begin{equation}\label{eq34}
\left(p\!\!\!/ - A \!\!\!/\right)^2=\left(p-A\right)+\frac 1 2
\sigma_{\mu\nu}\,F^{\mu\nu}\;\;\;
\left(F_{\mu\nu}=\partial_\mu A_\nu - \partial_\nu A_\mu\right)
\end{equation}
so that in accordance with (\ref{eq15}) we must compute
\begin{equation}\label{eq35}
\begin{array}{c}
{\rm tr}<x|exp-\left[
\displaystyle{\frac 1 2} \left(p-A\right)^2+
\displaystyle{\frac 1 4}\sigma_{\mu\nu}F^{\mu\nu}\right]\,t|y>
\\
\\
=
{\rm tr}\;\displaystyle{\sum_{n=-\infty}^{\infty}}{\rm e}^{i\pi n}\;
{\rm P} \!\!\!\!\!\:\! {\rm I}\;\;
\displaystyle{
\int_{\!\!\!\!\!\!\!\!\!{_{{_{q\left(0\right)=y}}}}}
^{\!\!\!\!\!\!\!\!{^{^{q\left(t\right)=\left(x_0+n\beta,\vec 
x\right)}}}}}
\!\!\!\!\!\!\!\!\!\!\!\!\!\!\!\!
{\cal D} q\left(\tau\right)
{\rm exp}\left\{\displaystyle{\int}_0^t {\rm d} \tau 
\left[\displaystyle{-\frac 1 2} \dot{q}^2\left(\tau\right)+
i A\left(q\left(\tau\right)\right)\cdot \dot{q}\left(\tau\right) - 
\displaystyle{\frac 1 
4}\sigma_{\alpha\beta}F^{\alpha\beta}\left(q\left(\tau
\right)\right)\right]\right\}
\end{array}.
\end{equation}
The term contributing to the path integral in (\ref{eq35}) that is of
fourth order in the background field $A_\mu$ is
\begin{equation}\label{eq36}
\begin{array}{c}
\Lambda={\rm tr}\;\displaystyle{\sum_{n=-\infty}^{\infty}}{\rm 
e}^{i\pi n}\;
\displaystyle{
\int_{\!\!\!\!\!\!\!\!\!{_{{_{q\left(0\right)=y}}}}}
^{\!\!\!\!\!\!\!\!{^{^{q\left(t\right)=\left(x_0+n\beta,\vec 
x\right)}}}}}
\!\!\!\!\!\!\!\!\!\!\!\!\!\!\!\!
{\cal D} q\left(\tau\right)
{\rm e}^{-\frac 1 2\int_0^t {\rm d} \tau \, 
\dot{q}^2\left(\tau\right)}
\int_0^t {\rm d} \tau_1  \int_0^{\tau_1} {\rm d} \tau_2 
\int_0^{\tau_2} {\rm d} \tau_3 \int_0^{\tau_3} {\rm d} \tau_4
\\
\left[4 \, \dot{q}_1\cdot A_1\,\dot{q}_2\cdot A_2
      \,\dot{q}_3\cdot A_3\,\dot{q}_4\cdot A_4-
\displaystyle{\frac 1 2}\left(F_{1_{\alpha\beta}}F_{2_{\alpha\beta}}
 \,\dot{q}_3\cdot A_3\,\dot{q}_4\cdot A_4+{\rm perm.}\right)
\right.
\\
\left.
-\displaystyle{\frac 1 2}
\left(F_{1_{\alpha\beta}}F_{2_{\beta\gamma}}F_{3_{\gamma\alpha}}
\,\dot{q}_4\cdot A_4+{\rm perm.}\right)
\right.
\\
\left.
+\displaystyle{\frac 1 4}\left(
F_{1_{\alpha\beta}}F_{2_{\beta\gamma}}
F_{3_{\gamma\delta}}F_{4_{\delta\alpha}}-
F_{1_{\alpha\beta}}F_{3_{\beta\gamma}}
F_{4_{\gamma\delta}}F_{2_{\delta\alpha}}-
F_{1_{\alpha\beta}}F_{4_{\beta\gamma}}
F_{2_{\gamma\delta}}F_{3_{\delta\alpha}}\right)
\right.
\\
\left.
+\displaystyle{\frac {1}{16}}\left(
F_{1_{\alpha\beta}}F_{2_{\alpha\beta}}
F_{3_{\gamma\delta}}F_{4_{\gamma\delta}}+
F_{1_{\alpha\beta}}F_{3_{\alpha\beta}}
F_{2_{\gamma\delta}}F_{4_{\gamma\delta}}+
F_{1_{\alpha\beta}}F_{4_{\alpha\beta}}
F_{2_{\gamma\delta}}F_{3_{\gamma\delta}}\right)\right].
\end{array}
\end{equation}
(We have let $q_i\equiv q\left(\tau_i\right)$ and 
$A_{\mu_i}\equiv A_\mu\left(q\left(\tau_i\right)\right)$, and 
``perm.''
refers to permutations of 1, 2, 3, 4. The trace identities
$$
\begin{array}{l}
{\rm tr}\;\sigma_{\alpha\beta}=0\\
{\rm tr}\;\sigma_{\alpha\beta}\;\sigma_{\gamma\delta}\;F_{\alpha\beta}
 \;G_{\gamma\delta}=-8\,{\rm tr}\left(F\,G\right)\\
{\rm 
tr}\;\sigma_{\alpha\beta}\;\sigma_{\gamma\delta}\;\sigma_{\lambda\rho}
\;F_{\alpha\beta}\;G_{\gamma\delta}\;H_{\lambda\rho}  
=-32 i\,{\rm tr}\;\left(F\,G\,H\right)\\
{\rm 
tr}\;\sigma_{\alpha\beta}\;\sigma_{\gamma\delta}\;\sigma_{\lambda\rho}
\;\sigma_{\mu\nu}
\;F_{\alpha\beta}\;G_{\gamma\delta}\;H_{\lambda\rho}\;I_{\mu\nu}=
64\;{\rm tr}\left(F\,G\,H\,I-F\,H\,I\,G-F\,I\,G\,H\right)\\
+16\left[{\rm tr}\left(F\,G\right)\;\left(H\,I\right)+
                 \left(F\,H\right)\;\left(G\,I\right)+
                 \left(F\,I\right)\;\left(G\,H\right)\right]
\end{array}
$$
have also been  used.) In order to evaluate the path integral in 
(\ref{eq36}), we use the plane wave expansion of (\ref{eq17}). The 
standard 
result of (\ref{eq22}) and the formula
\begin{equation}\label{eq37}
\sum_{n=-\infty}^{n=\infty}{\rm e}^{i\pi n - 
\left(n\beta\right)^2/2t}=
\frac{\sqrt{2\pi t}}{\beta} 
\sum_{n=-\infty}^{n=\infty}{\rm e}^{-
2\pi^2\,t\left(n+1/2\right)^2/\beta^2},
\end{equation}
together lead to the contribution of the $\zeta$-function to the four-
point
process to be
\begin{equation}\label{eq38}
\begin{array}{c}
\Gamma\left(s\right)\zeta^{(4)}\left(s\right)=
\int_0^\infty{\rm d}t t^{s-1}
\left(\displaystyle{\frac{1}{\left(2\pi\right)^{3/2}\beta^{1/2}}}\right)
\displaystyle{\sum_{n=-\infty}^{\infty}}
\displaystyle{\frac{\sqrt{2\pi\,t}}{\beta}}
{\rm e}^{-2\pi^2\,t\left(n+1/2\right)^2/\beta^2}\\
\times
\displaystyle{\int_0^t}{\rm d}\tau_1
\displaystyle{\int_0^{\tau_1}}{\rm d}\tau_2
\displaystyle{\int_0^{\tau_2}}{\rm d}\tau_3
\displaystyle{\int_0^{\tau_3}}{\rm d}\tau_4
\left[\displaystyle{\frac{\left(2\pi\right)^4\delta\left(\sum 
k_i\right)
{\rm exp}\left(\frac 1 2 k_i\cdot k_j \, G_{ij}\right)}
{\left(2\pi\, t\right)^2}}\right]\\
\times
\left\{4\left[\epsilon_1\cdot\epsilon_2\,\epsilon_3\cdot\epsilon_4\,
\ddot{G}_{12}\, \ddot{G}_{34}+{\rm perm.}\right.\right.\\
+ \left(\epsilon_1\cdot\epsilon_2\,\ddot{G}_{12}\right)
 \left(k_i\,{G}^{\dot{}}_{i3}\,\epsilon_3\right)
 \left(k_j\,{G}^{\dot{}}_{j4}\,\epsilon_4\right)+{\rm perm.}\\
\left.+ \left(k_i\,{G}^{\dot{}}_{i1}\,\epsilon_1\right)
 \left(k_j\,{G}^{\dot{}}_{j2}\,\epsilon_2\right)
 \left(k_l\,{G}^{\dot{}}_{l3}\,\epsilon_3\right)
 \left(k_m\,{G}^{\dot{}}_{m4}\,\epsilon_4\right) \right]\\ 
+\displaystyle{\frac 1 2}
\left[{\rm tr}\left(f_1\,f_2\,f_3\right)
\left(k_i\,{G}^{\dot{}}_{i4}\,\epsilon_4\right)+{\rm perm}\right]\\
+\displaystyle{\frac 1 4}
\left[{\rm tr}\left(f_1\,f_2\,f_3\,f_4-
                          f_1\,f_3\,f_4\,f_2-
                          f_1\,f_4\,f_2\,f_3\right)\right.\\
\left.
+\displaystyle{\frac{1}{16}}
\left({\rm tr}\left(f_1\,f_2\right)
      {\rm tr}\left(f_3\,f_4\right)+
      {\rm tr}\left(f_1\,f_3\right)
      {\rm tr}\left(f_4\,f_2\right)+
      {\rm tr}\left(f_1\,f_4\right)
      {\rm tr}\left(f_2\,f_3\right)\right)\right]
\end{array},
\end{equation}
where 
$f_{i\alpha\beta}\equiv-i\left(k_{i\alpha}\epsilon_{i\beta}-
                               k_{i\beta}\epsilon_{i\alpha}\right)$,
$G_{ij}=\frac 1 2|\tau_i-\tau_j|-\frac 1 2\left(\tau_i+\tau_j\right)+
\tau_i\tau_j/t$, 
${G}^{\dot{}}_{ij}=\frac{{\rm d}}{{\rm d}\tau_j}G_{ij}$ and
$\ddot{G}_{ij}=\frac{{\rm d}^2}{{\rm d}\tau_i {\rm d}\tau_j}G_{ij}$.
The rescaling $\tau_i=\sigma_i\,t$ allows us to integrate over $t$;
we can then sum over $n$ using \cite{ref13,ref14}
\begin{equation}\label{eq39}
\begin{array}{c}
\displaystyle{\sum_{n=-\infty}^{\infty}} 
\left[\nu^2 + \left(n + 
\displaystyle{\frac{\theta}{2\pi}}\right)^2\right]^{-s} =
\displaystyle{\frac{\sqrt{\pi}}{\Gamma(s)}} \left[
\displaystyle{\frac{\Gamma(s -\frac{1}{2})}{\nu^{2s-1}}} + 
4 \displaystyle{\sum_{m=1}^\infty} 
\cos(m \theta)\left(
\displaystyle{\frac{\pi m}{\nu}}\right)^{s-\frac{1}{2}}
\right.
\\
\left.
\times
K_{s-\frac{1}{2}} (2 \pi\,\nu\,m)\right]
\end{array},
\end{equation}
so that we are dealing with expressions of the form
\begin{equation}\label{eq40}
\begin{array}{c}
\displaystyle{\sum_{n=-\infty}^{\infty}} 
\Gamma\left(s+\displaystyle{\frac 3 2 }+\epsilon\right)\left[
\displaystyle{\frac{\Lambda^2}{2}}+
\displaystyle{\frac{2\pi^2}{\beta^2}}
\left(n+\displaystyle{\frac 1 2}\right)\right]^
{-\left(s+\frac 3 2+\epsilon\right)}\\
=\left(\displaystyle{\frac{2\pi^2}{\beta^2}}\right)^
{-\left(s+\frac 3 2+\epsilon\right)}\sqrt{\pi}\left\{
\displaystyle{\frac{\Gamma\left(s+1+\epsilon\right)}
{\left(\beta\Lambda/2\pi\right)^{2\,s+2+2\epsilon}}}+
4\displaystyle{\sum_{m=1}^{\infty}}\left(-1\right)^m
\left(\displaystyle{\frac{m\pi}{\beta\Lambda/2\pi}}\right)^{s+1
+\epsilon}\right.
\\
K_{s+1+\epsilon}\left(m\beta\Lambda\right)
\Biggr\}.
\end{array}
\end{equation}
(Here we have $\epsilon=-1,\,0,\,1$ and $\Lambda^2=-k_i\,g_{ij}\,k_j$,
with $g_{ij}\equiv\frac 1 t G\left(\sigma_i t, \sigma_j t\right)$.)
Using the integral representation\cite{ref15}
\begin{equation}\label{eq41}
K_a\left(z\right)=\int_0^\infty{\rm d}t\,{\rm cosh}\left(a\, t\right)
{\rm e}^{-z\, {\rm cosh}\left(t\right)},
\end{equation}
allows us to evaluate the sum over $m$ in (\ref{eq40}). The Mellin 
formula
\cite{ref14}
\begin{equation}\label{eq42}
\begin{array}{c}
\displaystyle{\frac{1}{{\rm e}^\lambda +1}}=
\displaystyle{\int_C}
\displaystyle{\frac{{\rm d}z}{2\pi i}}\zeta_{+}\left(z\right)
\Gamma\left(z\right)\,\lambda^{-z}
\\
\left( \zeta_{+}(z) \equiv \sum_{k=1}^\infty
\displaystyle{\frac{(-1)^{k+1}}{k^z}}=(1 - 2^{1-
z})\zeta\left(z\right);\;
C {\rm is\;the\;contour\;}\rm{Re}\;z>1-2\epsilon\right)
\end{array}
\end{equation}
can then be employed; by using it we can integrate over the parameter
$t$ used in (\ref{eq41}) since \cite{ref15}
\begin{equation}
\int_0^\infty \frac{{\rm d}t}{(\cosh t)^a} = \frac{\sqrt{\pi}}{2}
\frac{\Gamma(\frac{a}{2})}{\Gamma(\frac{a+1}{2})}.
\end{equation}
A final simplification involves using the formula \cite{ref15}
\begin{equation}\label{eq44}
\Gamma(2\,z)=\frac{2^{2z-1}}{\sqrt{\pi}} \Gamma(z)
\Gamma(z+\frac{1}{2});
\end{equation}
at last we get
\begin{equation}
\begin{array}{c}
\left.\displaystyle{\frac{{\rm d}}{{\rm d} s}}\right|_{s=0}
\displaystyle{\frac{\Gamma\left(s+3/2+\epsilon\right)}
{\Gamma\left(s\right)}}
\displaystyle{\sum_{n=-\infty}^{\infty}}
\left[
\displaystyle{\frac{\Lambda^2}{2}}+
\displaystyle{\frac{2\pi^2}{\beta^2}}\left(n+1/2\right)^2
\right]^{-\left(s+3/2+\epsilon\right)}
\\
\\
=\left\{
\begin{array}{lr}
\left(\displaystyle{\frac{2\pi^2}{\beta^2}}\right)^{-\frac 5 
2}\!\!\!\!\!
\sqrt{\pi}
\left[\left(\displaystyle{\frac{2\pi}{\beta\Lambda}}\right)^4+
4\left(\displaystyle{\frac{2\pi^2}{\beta\Lambda}}\right)^2
\displaystyle{\int_C}
\displaystyle{\frac{{\rm d}z}{2\pi i}}
\zeta_{+}\left(z\right)\,2^{z-2}\left(z\right)\left(z+2\right)
\Gamma^2\left(\frac z 2\right)\left(-\beta\Lambda\right)^{-z-2}\right]
& \left(\epsilon=1\right)
\\ \\
\left(\displaystyle{\frac{2\pi^2}{\beta^2}}\right)^{-\frac 3 
2}\!\!\!\!\!
\sqrt{\pi}
\left[\left(\displaystyle{\frac{2\pi}{\beta\Lambda}}\right)^2+
4\left(\displaystyle{\frac{-2\pi^2}{\beta\Lambda}}\right)
\displaystyle{\int_C}
\displaystyle{\frac{{\rm d}z}{2\pi i}}
\zeta_{+}\left(z\right)\,2^{z-2} z
\Gamma^2\left(\frac z 2\right)\left(-\beta\Lambda\right)^{-z-1}\right]
& \left(\epsilon=0\right)
\\ \\
\left(\displaystyle{\frac{2\pi^2}{\beta^2}}\right)^{-\frac 1 
2}\!\!\!\!\!
\sqrt{\pi}
\left[{\rm ln}\left(\displaystyle{\frac{2}{\Lambda^2}}\right)-
4 \displaystyle{\int_C}
\displaystyle{\frac{{\rm d}z}{2\pi i}}
\zeta_{+}\left(z\right)\,2^{z-2}
\Gamma^2\left(\frac z 2\right)\left(\beta\Lambda\right)^{-z}\right]
& \left(\epsilon=-1\right)
\end{array}\right.
\end{array}.
\end{equation}
With this result we see that (\ref{eq38}) becomes
\begin{equation}\label{eq46}
\begin{array}{c}
\zeta^{{(4)}'}(0)=\sqrt{\pi}
\displaystyle{\int_0^1}{\rm d}\sigma_1
\displaystyle{\int_0^{\sigma_1}}{\rm d}\sigma_2
\displaystyle{\int_0^{\sigma_2}}{\rm d}\sigma_3
\displaystyle{\int_0^{\sigma_3}}{\rm d}\sigma_4
\displaystyle{\frac{\delta\left(\sum k_i\right)}
{\left(2\pi\right)^{7/2}\beta^3}}
\\
\times
\left\{
\left(\displaystyle{\frac{2\pi^2}{\beta^2}}\right)^{-1/2}
\left[
{\rm ln}\left(\displaystyle{\frac{2}{\Lambda^2}}\right)-
4\displaystyle{\int_C}\displaystyle{\frac{{\rm d}z}{2\pi i}}
\zeta_{+}\left(z\right)\,2^{z-2}\Gamma\left(z/2\right)
\left(\beta\Lambda\right)^{-z}\right]\right.
\\
\left[
4\,\epsilon_1\cdot\epsilon_2\,\epsilon_3\cdot\epsilon_4
\left(-\delta\left(\sigma_1-\sigma_2\right)+1\right)
\left(-\delta\left(\sigma_3-\sigma_4\right)+1\right)
+{\rm perm.}\right]
\\ 
+\left(\displaystyle{\frac{2\pi^2}{\beta^2}}\right)^{-3/2}
\left[\left(\displaystyle{\frac{2\pi}{\beta\Lambda}}\right)^2+
4\left(\displaystyle{\frac{-2\pi^2}{\beta\Lambda}}\right)
\displaystyle{\int_C}
\displaystyle{\frac{{\rm d}z}{2\pi i}}
\zeta_{+}\left(z\right)\,2^{z-2} z
\Gamma^2\left(\frac z 2\right)\left(-\beta\Lambda\right)^{-z-1}\right]
\\
\left[
4\,\epsilon_1\cdot\epsilon_2\,
\left(-\delta\left(\sigma_1-\sigma_2\right)+1\right)
\displaystyle{\sum_i}k_i\cdot\epsilon_3
\left(-\displaystyle{\frac 1 2}\epsilon\left(\sigma_i-\sigma_3\right)-
       \displaystyle{\frac 1 2}+\sigma_i\right)
\displaystyle{\sum_j}k_j\cdot\epsilon_4
\left(-\displaystyle{\frac 1 2}\epsilon\left(\sigma_j-\sigma_4\right)-
       \displaystyle{\frac 1 2}+\sigma_j\right)\right.
\\
\left.
+ {\rm perm.}
-\displaystyle{\frac 1 2} f_{1\alpha\beta}\,f_{2\alpha\beta}\,
\epsilon_3\cdot\epsilon_4\left(-\delta\left(\sigma_3-
\sigma_4\right)+1\right)
+ {\rm perm.}\right]
\\
+
\left(\displaystyle{\frac{2\pi^2}{\beta^2}}\right)^{-\frac 5 
2}\!\!\!\!\!
\sqrt{\pi}
\left[\left(\displaystyle{\frac{2\pi}{\beta\Lambda}}\right)^4+
4\left(\displaystyle{\frac{2\pi^2}{\beta\Lambda}}\right)^2
\displaystyle{\int_C}
\displaystyle{\frac{{\rm d}z}{2\pi i}}
\zeta_{+}\left(z\right)\,2^{z-2}\left(z\right)\left(z+2\right)
\Gamma^2\left(\frac z 2\right)\left(-\beta\Lambda\right)^{-z-2}\right]
\\
\left[4
\displaystyle{\sum_i}k_i\cdot\epsilon_1
\left(-\displaystyle{\frac 1 2}\epsilon\left(\sigma_i-\sigma_1\right)-
       \displaystyle{\frac 1 2}+\sigma_i\right)\cdots
\displaystyle{\sum_j}k_j\cdot\epsilon_4
\left(-\displaystyle{\frac 1 2}\epsilon\left(\sigma_i-\sigma_4\right)-
       \displaystyle{\frac 1 2}+\sigma_i\right)
\right.
\\
+
\displaystyle{\frac 1 2}{\rm tr}f_1\,f_2
\displaystyle{\sum_i}k_i\cdot\epsilon_3
\left(-\displaystyle{\frac 1 2}\epsilon\left(\sigma_i-\sigma_3\right)-
       \displaystyle{\frac 1 2}+\sigma_i\right)
\displaystyle{\sum_j}k_j\cdot\epsilon_4
\left(-\displaystyle{\frac 1 2}\epsilon\left(\sigma_i-\sigma_4\right)-
       \displaystyle{\frac 1 2}+\sigma_i\right)
+{\rm perm}
\\
+
\displaystyle{\frac 1 2}{\rm tr}\left(f_1\,f_2\,f_3\right)
\displaystyle{\sum_j}k_j\cdot\epsilon_4
\left(-\displaystyle{\frac 1 2}\epsilon\left(\sigma_j-\sigma_4\right)-
       \displaystyle{\frac 1 2}+\sigma_j\right)+{\rm perm.}
\\
+
\displaystyle{\frac 1 4}
{\rm tr}\left(f_1\,f_2\,f_3\,f_4-f_1\,f_3\,f_4\,f_2-
f_1\,f_4\,f_2\,f_3\right)
\\
+
\left.\left.
\displaystyle{\frac{1}{16}}
\left(
{\rm tr}f_1f_2\,{\rm tr}f_3f_4+
{\rm tr}f_1f_3\,{\rm tr}f_2f_4+
{\rm tr}f_1f_4\,{\rm tr}f_2f_3
\right)
\right]
\right\}
\end{array}
\end{equation}
The integrals over $\sigma_i$ eliminates the first term in 
(\ref{eq46}).
Closing the contour $C$ with a semi-circle at infinity on the left 
side
of the complex plane allows us to evaluate the integrals over $z$ in
(\ref{eq46}) by use of the residue theorem. The sum over the
contributions of these residues gives rise to a power series in
$T^{-1}=\beta$, starting at $T^2$. This is consistent with \cite{ref7}
once we recall the dependence on $\beta$ in the normalization of the
wave function in (\ref{eq17}). There is no dependence on 
${\rm ln}\beta$ since 
$z\zeta_{+}\left(z\right)\Gamma^2\left(z/2\right)$
has only a single pole at $z=0,-2,-4,-6\cdots$; this is consistent 
with
\cite{ref7} as well.

In \cite{ref11} it is demonstrated that the vacuum polarization in
thermal quantum electrodynamics is transverse; it is anticipated that 
one
can similarly show that the four-point function in (\ref{eq46}) is 
also
gauge invariant.

\section{Discussion}
In this paper we have demonstrated how thermal effects in field 
theory can
be computed perturbatively in the loop expansion by using either a so-
called
``Barton Expansion'' or the quantum mechanical path integral. The two
approaches are significantly different in their technical details;
nevertheless we have demonstrated that the results obtained in 
\cite{ref7} for the four point function at one loop order in thermal
QED using the ``Barton expansion'' are consistent with what can be 
obtained
using QMPI.

\acknowledgements{D. G. C. M. would like to thank FAPESP (Brazil) for
financial support and the Universidade de S\~ao Paulo for its 
hospitality
while this work was being conducted. F. T. B. is grateful to 
CNPq (Brazil) for a grant.}

\end{document}